\documentclass{ws-mpla}

\begin{document}
\markboth{R.\,R.\,Abbyazov, S.\,V.\,Chervon} {Unified dark matter
and dark energy description in a chiral cosmological model}
\catchline{}{}{}{}{}

\title{UNIFIED DARK MATTER AND DARK ENERGY
DESCRIPTION \\IN A CHIRAL COSMOLOGICAL MODEL\\
}
\author{\footnotesize RENAT R. ABBYAZOV
}

\address{Department of Physics, Ulyanovsk State Pedagogical University
named after I.N. Ulyanov,\\
100 years V.I. Lenin's Birthday Square, 4, 432700 Ulyanovsk,
Russia
\\
renren2007@yandex.ru}

\author{\footnotesize SERGEY V. CHERVON
}
\address{Astrophysics and Cosmology Research Unit \\
School of Mathematics, Statistics and Computer Science, University of KwaZulu-Natal \\
Private Bag X54 001, Durban 4000, South Africa \footnote{The
permanent address: Department of Physics, Ulyanovsk State
Pedagogical University named after I.N. Ulyanov, 100 years V.I.
Lenin's Birthday Square,
4, 432700 Ulyanovsk, Russia}\\
chervon.sergey@gmail.com }

\maketitle

\pub{Received (Day Month Year)}{Revised (Day Month Year)}

\begin{abstract}

We show the way of dark matter and dark energy presentation via ansatzs on the kinetic energies of the fields in the two-component chiral cosmological model. To connect a kinetic interaction of dark matter and dark energy with observational data the reconstruction procedure for the chiral metric component $h_{22}$ and the potential of (self)interaction $V$ has been developed. The reconstruction of $h_{22}$ and $V$ for the early and later inflation have been performed. The proposed model is confronted to $\Lambda CDM$ model as well.

\end{abstract}

\ccode{PACS Nos.: 98.80.-k, 95.36.+x}

\section{Introduction}
\numberwithin{equation}{section}
\newcommand{\ph}{\varphi}
\newcommand{\prt}{\partial}

The later-time cosmic acceleration of our Universe is strongly
supported by observational data. Namely observations of supernovae
type Ia\cite{Suzuki:2011hu}, the data from Baryon Acoustic
Oscillations (BAO)\cite{Percival:2009xn} and Cosmic Microwave
Background (CMB)\cite{Komatsu:2010fb} measurements confirm that the
Universe is expending with an acceleration at the present time and
about 70\% of the energy density consists of dark energy in a wide
sense\cite{tsuji-10}, i.e. as the substance which is responsible for an
anti-gravity force.

In the range with well--known $\Lambda$CDM model, which
potentially provides correct description of the Universe evolution
but suffers from fine--tuning and coincidence problems, some
alternative models were proposed. We will pay attention to the
models with presence of scalar fields included in quintessence,
phantom and quintom \cite{cosats06,DarkEnergy1103,Padmanabhan,QuintomReview}
models.

A chiral cosmological model (CCM) as a nonlinear sigma model with
a potential of (self)interactions\cite{chervon12qm} has been already used extensively in various areas
of gravitation and cosmology\cite{ch97gc,ch02gc,brochesush}
and in particular for description of the very early
Universe\cite{bcmk12qm,Beesham} and
inflation\cite{chzhsh97plb,chekos2003}.
A CCM can be applicable as well to the late-time Universe with
dark matter and dark energy domination as it was shown in
\cite{panche11}.

The purpose of this article is to put into use the two-component CCM as
the model where the dark energy content of the Universe and also
the dark matter component are represented by two chiral fields with
kinetic and potential interactions\cite{chervon12qm}.
By considering a target space metric in the form 
\begin{equation}\label{tsm}
    ds_{\sigma}^2=h_{11}d\varphi^2 +h_{22}(\varphi,\chi)d\chi^2,~~~h_{11}=const.
\end{equation}
we prescribe a kinetic interaction between chiral fields $\varphi$
and $\chi $ as a functional dependence $h_{22}$ on the fields. The
potential interaction will be included into standard potential
energy term of the action.

There are no enough indications from observations about kinetic
interactions between dark sector fields. Therefore we always deal
with the problem: what is the functional dependence for the chiral
metric component on the fields?
First idea is to attract some results from HEP, for example, to
consider SO(3) symmetry (by taking $h_{22}=\sin^2 \varphi $)
and/or others symmetries for a chiral space. 
From the other hand one can use some testing kinetic
interactions\cite{brochesush,panche11}.

Thus we can state that there is no evidence for some preferable
functional form of the kinetic interaction contained in the
functional form of the $h_{22}$ chiral metric component.
To avoid this problem we develop here 
the reconstruction procedure for the chiral metric component
$h_{22}$.
We ascribe a certain desirable behavior on the kinetic energy of
the second chiral field $\chi $ and it becomes possible to
determine both the target space metric component $h_{22}$ and a
(self)interacting potential $V$ depending on the first chiral
field $\varphi $.
So we can restore a
functional dependence the $h_{22}$ and $V$ on the scalar field $\varphi $ using observational data.
Unfortunately it turns out that the procedure could not be applied
for the entirely Universe evolution and we have necessity to consider
separately the  early and late epochs of the Universe evolution.

It will be shown also that a CCM describes dark energy and
dark matter in the unified form under special restrictions on the
chiral fields (ansatzs).
Therefore to include into consideration the present Universe with
accelerated expansion it needs to take into account baryonic
matter and radiation in the range with a two-component CCM.

Making confrontation of proposed model predictions with
observational data we found the way of a reconstruction of a kinetic
interaction term $h_{22}$ and the potential $V$ in an exact form.
This reconstruction is based on the procedure of finding the
best--fit values 
matching to the astrophysical observations.

The structure of the article is like follow.  In section 2, we give
the basic model equations and discuss their properties including
the exact solutions for a pure CCM (without matter
and radiation). We derive the Friedmann equation for the
proposed model with the aim to make comparison with $\Lambda$CDM in section 3.
In section 4, we give the details of a fitting procedure outline.
We present the way of the reconstruction of the kinetic coupling and
potential in section 4. The early and recent Universe
approximations are discussed there as well. Section 6 is devoted to
the background dynamics of a CCM.
%
Finally in section 7, we discuss the obtained results and consider
perspectives for the future investigations.


\section{The model equations and their properties}
\numberwithin{equation}{section}

Recently we proposed a CCM coupling to a perfect
fluid\cite{arrche12gc} with the aim to investigate chiral fields
interaction with CDM. For the sake of shortness
we termed this model as $\sigma CDM$ to stress its difference from $\Lambda CDM, QCDM $ and others models.
$\sigma CDM$ model presents a generalization of a single scalar field
model coupled to CDM in the form of a perfect fluid\cite{Sur:2009jg}.
The model is described by the action functional

\begin{equation}\label{0}
    S=
    \int d^4x\sqrt{-g}\left(-\frac{1}{2}g^{\mu\nu}h_{AB}\prt_\mu\ph^A\prt_\nu\ph^B-
    V(\ph^C)\right)+S_{(pf)}.
\end{equation}
Here $S_{(pf)}$ stands for the perfect fluid part of the action,
$h_{AB}=h_{AB}(\varphi^C)$ are the target space metric components
depending on the scalar fields $\varphi^C $.
The line element of a target (chiral) space is
\begin{equation}\label{nsm}
    ds_{\sigma}^2=h_{AB}(\varphi^C)d\varphi^A d\varphi^B.
\end{equation}
We use shortened notations for the partial derivatives with respect
to the space-time
coordinates: $\frac{\partial \varphi^A}{\partial x^\mu} =
\prt_\mu\ph^A$. As usual $g_{\mu\nu}(x^\alpha) $ denotes a
space-time metric as a function on the space-time coordinates, so
Greek indices $\alpha, \mu, ... $ vary in a range from 0 to 3,
Latin capital letters $A$, $B, ...$ -- take values from 1 to $N$
where $N$ is evidently corresponding to the chiral fields number.

The space-time of homogeneous and isotropic Universe  is described
by a spatially-flat Friedmann -- Robertson -- Walker (FRW) metric
\begin{equation}\label{6}
    ds^2=-dt^2+a^2(t)\left(dr^2+r^2\left(d\theta^2+\sin^2\theta d\phi^2\right)\right).
\end{equation}
The two-component CCM has a target space metric simplified
to\cite{arrche12gc}
\begin{equation}\label{7}
    ds_{\sigma}^2=h_{11}d\varphi^2 +h_{22}(\varphi)d\chi^2,~~~h_{11}=const.
\end{equation}

The $\sigma CDM $ (\ref{0}) with internal space metric (\ref{7}) 
includes the models proposed earlier:
cold dark matter and cosmological constant ($\Lambda$CDM, when $h_{11}=h_{22}=0,~ V=const =\Lambda $)
model\cite{cosats06,DarkEnergy1103}, quintessence model (QCDM, when
$h_{11}=1, ~h_{22}=0$), phantom model (PhCDM, when $h_{11}=-1,~ h_{22}=0$),
quintom model (qCDM, when $h_{11}=1,~h_{22}=-1$)
\cite{QuintomReview,Generalizationquintom,Quintommixed,twoscalarfields}.
Thus the model under consideration is a
generalization of the models investigated earlier and mentioned
above.

As a first step of our study we consider the system of equations
of the two-component CCM without a perfect fluid. Using
assumptions $h_{11}=const$ and $h_{22}=h_{22}(\ph)$ expressed in
(\ref{7}) one can obtain the system of Einstein and chiral field
equations

\begin{align}
\label{1eq}
&H^2=\frac{8\pi G}{3}\left[\frac{1}{2}h_{11}\dot\ph^2+\frac{1}{2}h_{22}\dot\chi^2+V(\ph,\chi)\right],\\
\label{2eq}
&\dot H = -8\pi G\left[\frac{1}{2}h_{11}\dot\ph^2+\frac{1}{2}h_{22}\dot\chi^2\right],\\
\label{3eq}
&\ddot\ph+3H\dot\ph-\frac{1}{2h_{11}}\frac{d h_{22}}{d\ph}\dot\chi^2+\frac{1}{h_{11}}\frac{\prt V}{\prt\ph}=0,\\
\label{4eq}
&\ddot\chi+3H\dot\chi+\frac{1}{h_{22}}\frac{dh_{22}}{d\ph}\dot\ph\dot\chi+\frac{1}{h_{22}}\frac{\prt V}{\prt\chi}=0.
\end{align}
Here $H=\frac{\dot{a}}{a},~~(\dot{})=\frac{d}{dt}$.

When first inflationary models were analyzed it was much
attention to a very simple case when an inflationary potential
$V(\phi)$ equals to the constant\cite{Linde90}. Moreover this regime is very
important because it leads to an exponential expansion of the
Universe. Note that a scalar field is equal to a constant value as
well in this regime.

Let us consider for a minute the case of $V=const$ for the model
under consideration (\ref{1eq})-(\ref{4eq}). From (\ref{4eq}) one
can obtain\cite{chekos2003,chervon95aa}

\begin{equation}
  \label{eq:chi_sol_h22}
  \dot\chi^{2}=\frac{2C}{h_{22}^{2}a^{6}}.
\end{equation}

Combining (\ref{1eq}) and (\ref{2eq}) one can obtain the
well-known solution of a de Sitter Universe with Hubble parameter
and scale factor\cite{chervon12qm}
\begin{equation}\label{H-th}
H =\sqrt{\frac{\Lambda}{3}}\tanh (\sqrt{3\Lambda}t),~  a=a_*[\cosh
(\sqrt{3\Lambda}t)]^{1/3}.
\end{equation}
This solution with some approximation corresponds to the
inflationary stage of the Universe evolution. But our intention is
to proceed further in time therefore we need to include into consideration
radiation and matter to describe the present epoch of the Universe.

The method of the exact solutions construction for a CCM
(\ref{1eq})-(\ref{4eq}) is based on exploiting an additional
degree of freedom 
(see, for ex. discussion in\cite{Beesham}).
Namely even we fix the potential $V(\phi,\chi)$ there are still
four equations with four unknown functions
$H,~\varphi,~\chi,~h_{22}$ 
($h_{11}$ can be set
equal to $\pm 1$ without the loss of generality\cite{chervon12qm}).
Nevertheless the equation (\ref{1eq}) can be obtained from the
linear combination of the chiral field equations
(\ref{3eq})-(\ref{4eq}), so the equation (\ref{1eq}) doesn't
independent one. Therefore one may insert the symmetry on the
target space or can suggest a testing interaction between
chiral (dark sector) fields\cite{brochesush,chepan2010}.
Essentially new approach to this issue we propose here as
a reconstruction both $h_{22}$ and $V$ from observational data.

Let us
remind that for the scalar field cosmology by introducing
the selfinteracting potential  $V(\phi)$ we
have two equations with two unknown functions. (The same situation
will be if we set the dependence a scalar field on time or if we
know the scale factor of the Universe as a function on
time\cite{Beesham}).

To solve the system of a CCM interacting with a
perfect fluid (or matter) in explicit form is a very difficult
task. Therefore we will use an additional freedom connecting with
the chiral metric components $h_{11}$ and $h_{22}$ as a part of a
kinetic energy.

An interesting approach for a two-fields model with a cross
interaction was proposed in the work\cite{Sur:2009jg}. To describe
a dark matter component it was constructed
the special ansatzs for
the time derivatives of the scalar fields.
In our approach we will use instead some constraints on the kinetic parts of the
chiral fields (ansatzs) to obtain a correct description of the
present Universe.

Now let us turn our attention to a study of the model equations
(\ref{1eq})-(\ref{4eq}). It is easy to check that the solution 
(\ref{eq:chi_sol_h22}) for the constant potential will be valid
for the case when $V=V(\ph)$ only.
By extracting from  (\ref{eq:chi_sol_h22}) the kinetic energy term
for the field $\chi $ one can obtain
\begin{equation}\label{anz-chi}
\frac{1}{2}h_{22}\dot\chi^2=\frac{C}{h_{22}a^6}.
\end{equation}
We can ascribe by suggestion $h_{22}\sim a^{-3}$ dust matter like
behavior to the kinetic energy of the field $\chi$.
Using the behavior $h_{22}\sim a^{-3}$ it is easy to see that the
second field can be related to the dark matter term
provided the restriction to the kinetic energy of the second field
$\chi$ (ansatz)
\begin{equation}\label{anz-chi-a}
\frac{1}{2}h_{22}\dot\chi^2=Ca^{-3}.
\end{equation}
Let us mention here, that more simple ansatz
$\frac{1}{2}h_{22}\dot\chi^2=\Lambda_\psi=const $ has been
analyzed in\cite{chepan2008} and gave possibility to obtain the
exact solutions for the two-component CCM.
For the kinetic energy of the first field $\varphi$ we can form the
ansatz by a simple way
\begin{equation}\label{anz-vp}
\frac{1}{2}h_{11}\dot\ph^2=B=const.
\end{equation}
Further we will show that this relation is associated with the
dark energy component in the present Universe.

For convenience let us represent the ansatzs (\ref{anz-chi}), (\ref{anz-vp}) in the general forms:
\begin{eqnarray}
    \label{eq:f_g_ansatz-f}
\frac{1}{2}h_{11}\dot\ph^2=f(a),\\
\label{eq:f_g_ansatz}
  \frac{1}{2}h_{22}\dot\chi^2=g(a).
\end{eqnarray}
Thus we have $f(a)=B=const,~g(a)=Ca^{-3}$ in
(\ref{eq:f_g_ansatz-f})-(\ref{eq:f_g_ansatz}) and the chiral
metric component
\begin{equation}
  \label{eq:h22}
  h_{22}=a^{-3}.
\end{equation}

Let us note that the suggested restrictions above give rise to the
exact solution for the CCM describing by equations
(\ref{1eq})-(\ref{4eq}). Indeed from ansatzs we can find the
solutions for the chiral fields
\begin{equation}\label{sol-vp-chi}
\varphi =\sqrt{\frac{2B}{h_{11}}}t +\varphi_0,
~~\chi=\sqrt{2C}t+\chi_0.
\end{equation}
Then from Einstein equations (\ref{1eq})-(\ref{2eq}) we can define the
potential
\begin{equation}\label{V_sfa}
 V(a)=-6B\ln a +Ca^{-3}+V_* .
\end{equation}
The solution for the scale factor can be obtained from the
equation

\[ H^2=\frac{C_*}{a^6}+2\kappa \left( \frac{B}{6}-B\ln a +
\frac{C}{3a^3}+\frac{V_*}{6}\right). \]

It is difficult to find the scale factor in exact view from this general
equation, but for the special case assuming  $C_*=0$ and $C=0$
(under this assumption the second field $\chi $ becomes a constant),
we found that the Universe is in the stage with an exponential
expansion with $ a \propto \exp (Bt^2)$.

\section{A CCM coupling to barion matter and radiation. \\
Friedmann equation of the model}
\numberwithin{equation}{section}

Our following task is to connect the energy densities of various species
of the Universe to the Hubble parameter. To this end we need to include
into Friedmann equation (\ref{1eq}) the energy density of barion matter $ \rho_b$
and radiation $ \rho_r$. Thus (\ref{1eq}) for the recent Universe takes the form
\begin{equation}\label{fr-later}
  H^2=\frac{8\pi G}{3}\left[\rho_{\sigma}+\rho_{b}+\rho_{r}\right]
\end{equation}
where $ \rho_\sigma= \frac{1}{2}h_{11}\dot{\varphi}^2+ \frac{1}{2}h_{22}\dot{\chi}^2
+V$. Introducing the "pressure" of chiral fields $ p_\sigma= \frac{1}{2}h_{11}\dot{\varphi}^2+ \frac{1}{2}h_{22}\dot{\chi}^2
-V$ and using ansatzs (\ref{eq:f_g_ansatz-f})
and (\ref{eq:f_g_ansatz}) we can obtain
\[\rho_{\sigma}=f+g+V,\quad p_{\sigma}=f+g-V. \]

Using (\ref{V_sfa}) and extracting
the cosmological parameter $\Lambda$ from $V_*$ the energy density,
potential and pressure of the
two-component CCM can be expressed as
\begin{equation}
  \label{eq:rho_sigma}
\rho_{\sigma}=\Lambda-6B\ln a+2Ca^{-3},\quad\Lambda=B+V_*
\end{equation}
\begin{equation}
  \label{eq:pot_scal}
  V=\Lambda-6B\ln a+Ca^{-3}-B,
\end{equation}
\begin{equation}
  \label{eq:press_sigma}
  p_{\sigma}=2B-\Lambda+6B\ln a.
\end{equation}

By standard way (see, for ex. \cite{Garcia-Bellido}) one can define a critical density
$ \rho_c=\frac{3H_0^2}{8\pi G},$ where $H_0$ is the Hubble parameter of today expansion
 $ H_0=\frac{\dot{a}}{a}(t_0).$ Herefrom the subscript "0" is related to the present time $t_0$ when the scale factor $a(t_0)=a_0=1.$ Also we will use the density parameter
 $ \Omega_0=\frac{\rho}{\rho_c}(t_0)$ and the individual rations $ \Omega_i=\frac{\rho_i}{\rho_c}(t_0)$ for chiral fields, barion matter and radiation.

Let us remember that equations of
state for radiation and baryons are
\begin{equation*}
  p_{r}=\frac{1}{3}\rho_{r},\quad p_{b}=0,.
\end{equation*}
The energy densities and the contribution to the critical density can be represented as
\begin{equation*}
  \rho_{r}=\rho_{r0}a^{-4}=
\Omega_{r0}\rho_{c0}a^{-4},\quad\rho_{b}=\rho_{b0}a^{-3}=\Omega_{b0}\rho_{c0}a^{-3},\quad \rho_{c0}=\frac{3H^{2}_{0}}{8\pi G}.
\end{equation*}

Taking into account (\ref{eq:rho_sigma}) Friedmann equation (\ref{fr-later})
can be transformed to the normalised Hubble parameter form
\begin{equation*}
  \frac{H^2}{H_0^2}=\frac{1}{\rho_c}\left(\Lambda-6B\ln a+2Ca^{-3}\right)+\Omega_{b0}a^{-3}+\Omega_{r0}a^{-4}.
\end{equation*}
Making renormalization of the constants we finally obtain the
normalised Hubble rate in the form which is suitable for further
confronting with observational data
\begin{equation}
  \tilde{H^{2}}=\frac{H^2}{H_0^2}=\tilde\Lambda-6\tilde B\ln a+2\tilde Ca^{-3}+\Omega_{b0}a^{-3}+\Omega_{r0}a^{-4},
\end{equation}
where
\begin{equation}
  \tilde B=\frac{B}{\rho_c},\quad\tilde C=\frac{C}{\rho_c},\quad\tilde\Lambda=\frac{\Lambda}{\rho_c},\quad\tilde{H^{2}}=\frac{H^2}{H_0^2}.
\end{equation}

We need to find $\tilde\Lambda$ at $a=a_0=1$ with the help of
Friedmann equation. Cold dark matter (CDM) is included in the
model as the kinetic ansatz (\ref{anz-chi})
\begin{equation}
  \label{eq:sm_const_tilde} \tilde\Lambda=1-2\tilde
C-\Omega_{b0}-\Omega_{r0}=\Omega_{\sigma\Lambda0},\quad\Omega_{\sigma
cdm0}=2\tilde C,\quad\Omega_{m0}=\Omega_{\sigma cdm0}+\Omega_{b0}.
\end{equation}
Summing up the notations above
we display the final form of the normalised Hubble parameter
\begin{equation}
  \label{eq:Hubble_sigma_DE}
  \tilde{H^{2}}=\Omega_{\sigma\Lambda0}-6\tilde B\ln a+\Omega_{\sigma cdm0}a^{-3}+\Omega_{b0}a^{-3}+\Omega_{r0}a^{-4}.
\end{equation}

We propose here the $\sigma$CDM model containing the dark energy with variable equation of state. The model is an alternative to
$\Lambda\text{CDM}$ model with the cosmological constant and CDM.
Let us note that generally speaking the presence of
$\tilde{B}$ in (\ref{eq:Hubble_sigma_DE}) may change the values of $\Omega_{m0}$ and $\Omega_{\Lambda0}$, thus they can be distinctive from the corresponding quantities in
$\Lambda\text{CDM}$ model.
Nevertheless to find the exact values of this distinction we need to perform comparison with the experimental data.

\section{Comparison with experimental data }
\numberwithin{equation}{section}

From the very beginning\cite{Perlmutter}, \cite{Riess} supernovae Ia
type observations directly indicated an accelerated expansion of
the Universe. Observing supernovae luminosity distance $d_L$ as a
function of a redshift one can infer about an expansion history of the
Universe. Here we use one of the most recent compilation of the
supernovae sets Union 2.1 \cite{Suzuki:2011hu}. The procedure of
confronting cosmological model predictions with observations
consists of minimizing quantity procedure and calculation as
a result best--fit values of the model parameters for

\begin{equation}
  \chi^2_{SN}=\sum_{i=1}^{N}\frac{[\mu_{obs}(z_i)-\mu(z_i)]^2}{\sigma_i^2(z_i)}.
\end{equation}
Here as usual in the supernovae experimental analysis
\cite{Comparisondarkenergy} module distance $\mu(z_i)$ is used. The
dependence on the luminosity distance is

\begin{equation}
  \mu(z_i)=5\log_{10}[D_L(z_i)]+\mu_0,\quad D_L=H_0d_L,\quad \tilde H=H/H_0.
\end{equation}

\begin{equation}
  \mu_0=5\log_{10}\left[\frac{H_0^{-1}}{Mpc}\right]+25=42.38-5\log_{10}h,\quad H_0=\frac{h}{2998}\text{Mpc}^{-1}.
\end{equation}

In order to find more accurate parameter values and to reduce
errors significantly it is necessary to supplement the supernovae
observations with information about baryonic acoustic oscillations
(BAO) \cite{Percival:2009xn} and cosmic microwave background (CMB)
\cite{Komatsu:2010fb}.

BAO $\chi^2$ function is defined as

\begin{equation}
  \chi^2_{BAO}=\left(\frac{D_V(z=0.35)/D_V(z=0.2)-1.736}{0.065}\right)^2,
\end{equation}
where
\begin{equation}
  D_V\equiv\left[(1+z)^2D_A^2(z)\frac{z}{H(z)}\right]^{1/3}
\end{equation}
~is an effective distance measure, while
\begin{equation}
  D_A=(1+z)^{-2}d_L(z)
\end{equation}
is the angular diameter distance \cite{Percival:2009xn}.

Function $\chi^{2}$ for CMB is
\begin{equation}
  \chi^2_{CMB}=(x_i^{th}-x_i^{obs})(C^{-1})_{ij}(x_j^{th}-x_j^{obs}),
\end{equation}
where $x_i=(l_A,R,z_\ast)$~--- the vector of quantities which characterizes
the cosmological model and $(C^{-1})_{ij}$~--- WMAP7 covariance matrix
\cite{Komatsu:2010fb}. Here we use acoustic scale, from which
first acoustic peak of CMB power spectrum is depending on
\cite{tsuji-10}
\begin{equation}
  l_A\equiv(1+z_{\ast})\frac{\pi D_A(z_\ast)}{r_s(z_\ast)},
\end{equation}
which has been taken at the moment $z_{\ast}$ of decoupling of
radiation from matter, and on the sound horizon

\begin{equation}
  r_s(z)=\frac{1}{\sqrt{3}}\int^{1/(1+z)}_0\frac{da}{a^2H(a)\sqrt{1+(3\Omega_b/4\Omega_{\gamma})a}}.
\end{equation}

We will use the fitting formula

\begin{equation}
  z_\ast=1048[1+0.00124(\Omega_bh^2)^{-0.738}][1+g_1(\Omega_mh^2)^{g_2}],
\end{equation}
\begin{equation}
  g_1=\frac{0.0783(\Omega_bh^2)^{-0.238}}{1+39.5(\Omega_bh^2)^{0.763}},\quad
    g_2=\frac{0.560}{1+21.1(\Omega_bh^2)^{1.81}},
\end{equation}
for decoupling moment \cite{Hu:1995en}. 
Shift parameter $R$ is defined as 

\begin{equation}
  R(z_{\ast})=\sqrt{\Omega_{m0}H_{0}^{2}}(1+z_{\ast})D_{A}(z_{\ast}).
\end{equation}

Minimizing the sum
$\chi^{2}_{joint}=\chi^{2}_{SN}+\chi^2_{BAO}+\chi_{CMB}^2$ one can
find the best--fit $\tilde{B}$ and $\tilde{C}$ values. We also keep
fixed the radiation and baryonic contributions to the critical density
today $\Omega_{\gamma0}=2.469\cdot10^{-5}h^{-2}$,
$\Omega_{b0}=0.022765\cdot10^{-2}$, $h=0.742$. Also we take into
account a relativistic neutrino in addition to the photon radiation
component $\Omega_{r0}=(1+N_{eff})\Omega_{\gamma0}$, where
$N_{eff}=3.04$~--- the effective neutrino number
\cite{Comparisondarkenergy}. Our results for the best--fit from
$\chi^2_{joint}$ minimization are $\tilde{B}=0.00078, \Omega_{\sigma
m0}=\Omega_{b0}+2\tilde{C}=\Omega_{b0}+\Omega_{\sigma
cdm0}=0.23398$. For $\Lambda$CDM model we take best--fit values
$\Omega_{m0}=0.27$ and
$\Omega_{\Lambda0}=1-\Omega_{m0}-\Omega_{r0}$. To avoid confusion
between $\Omega_{cdm}$ and $\Omega_m$ in $\sigma$CDM and
$\Lambda\text{CDM}$ models we put additional index $\sigma$ in
$\Omega$ above.

\begin{figure}[h]
\center{\includegraphics[width=1\linewidth]{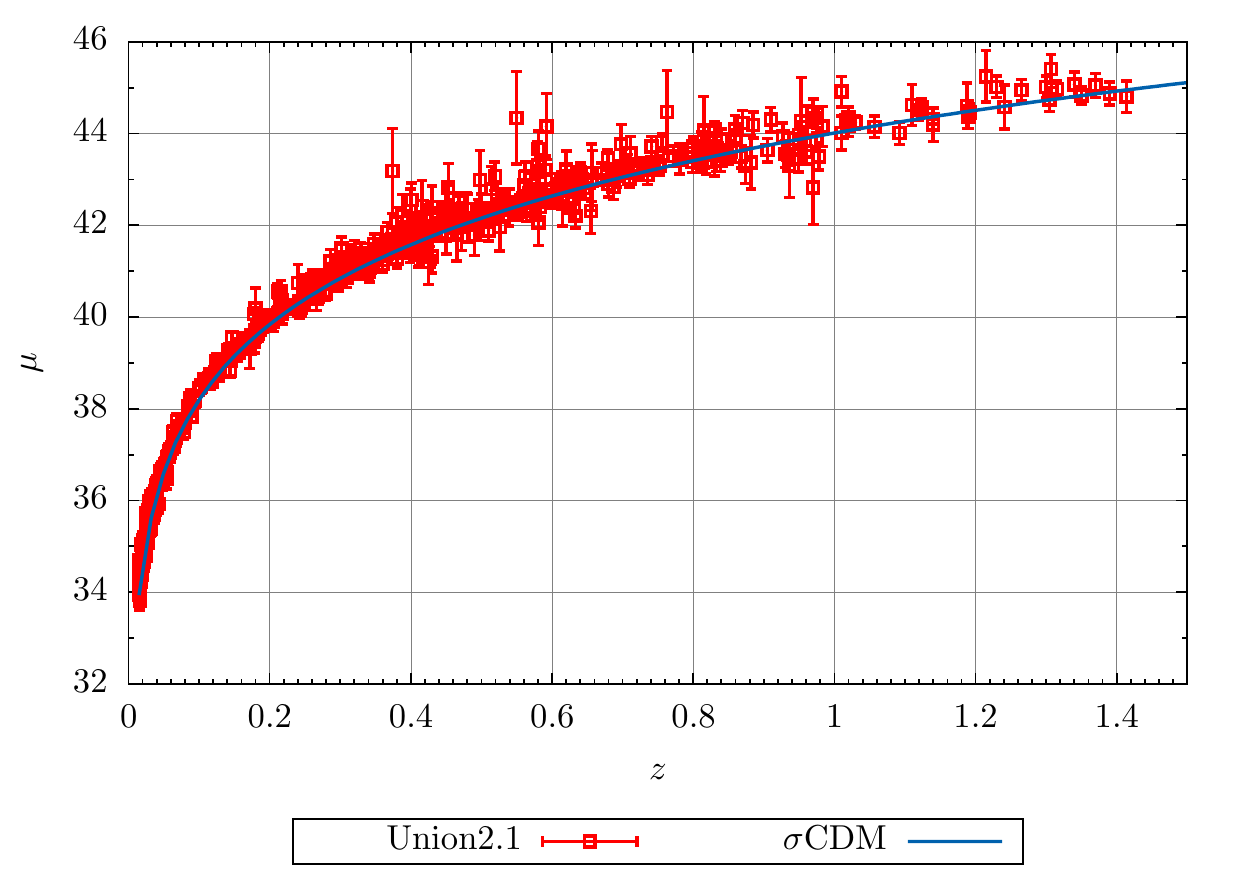}}
\caption{Supernovae Union 2.1 data and prediction from $\sigma
CDM$ model.} \label{ris:mu}
\end{figure}

\begin{figure}[h]
\center{\includegraphics[width=1\linewidth]{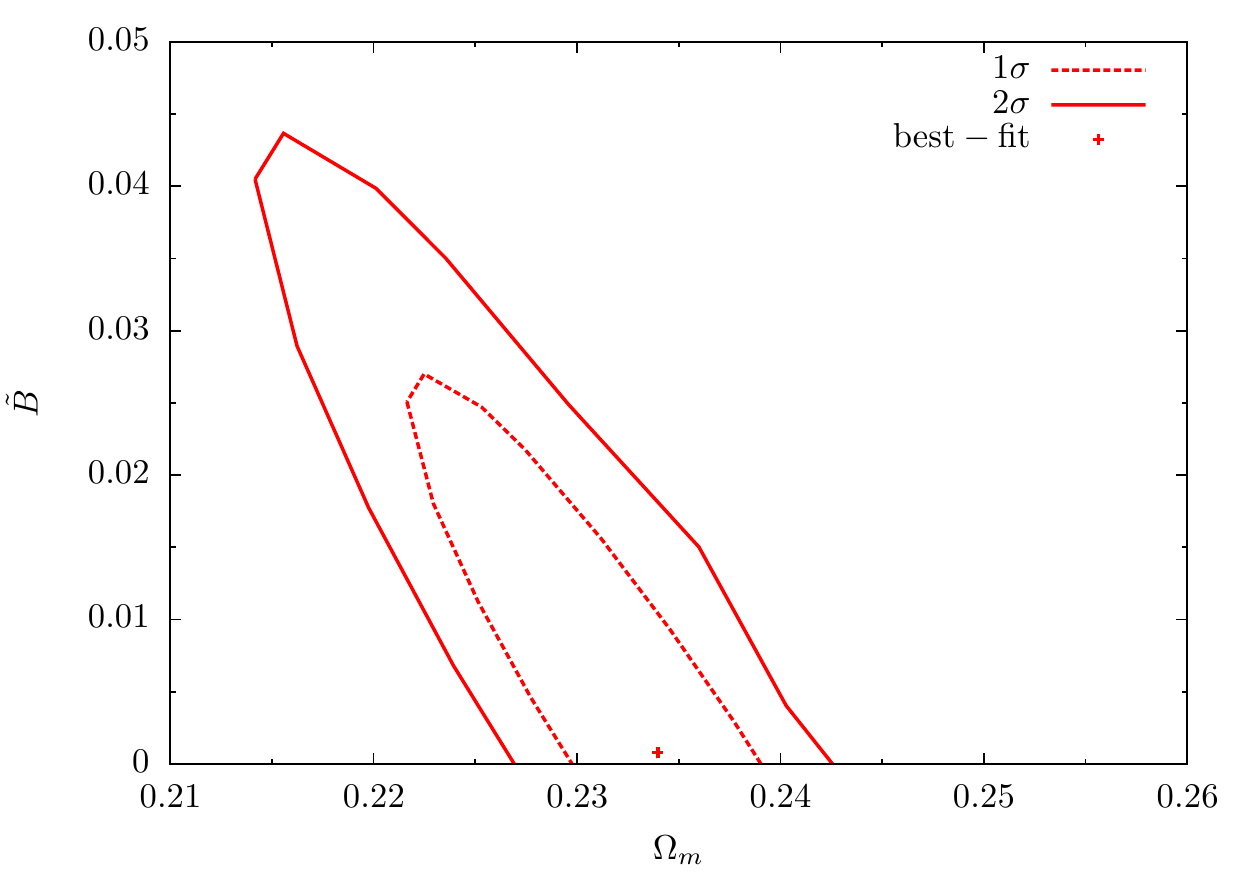}}
\caption{Contour plots corresponding to $1\sigma$ (68$\%$) $1\sigma$
and $2\sigma$ (95$\%$) likelihood levels for $\sigma\mathrm{CDM}$ model parameters.}
\label{ris:contours}
\end{figure}

\begin{figure}[h]
\center{\includegraphics[width=1\linewidth]{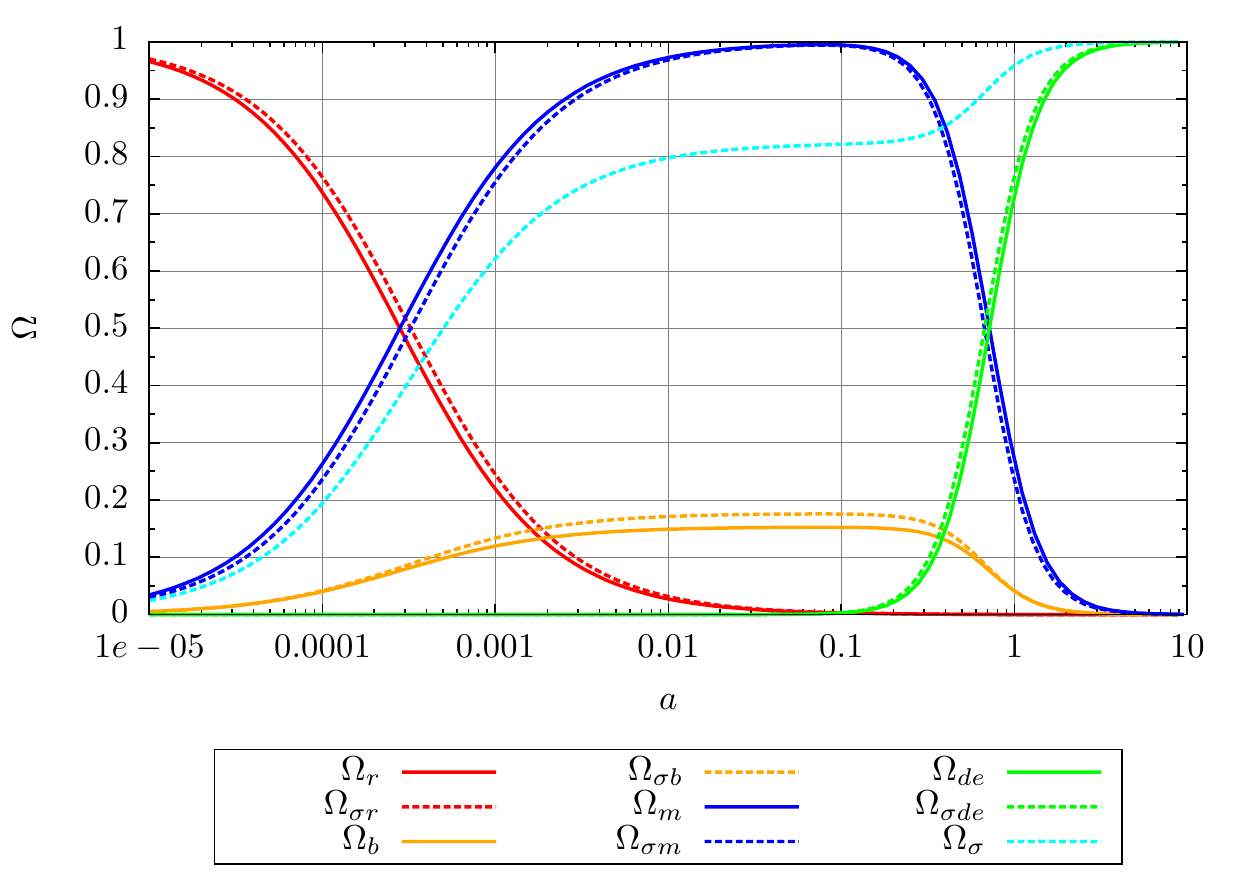}}
\caption{Evolution of contributions to critical density of various components in $\Lambda\mathrm{CDM}$ and $\sigma CDM$ models}
\label{ris:Omega}
\end{figure}

\begin{figure}[h]
\center{\includegraphics[width=1\linewidth]{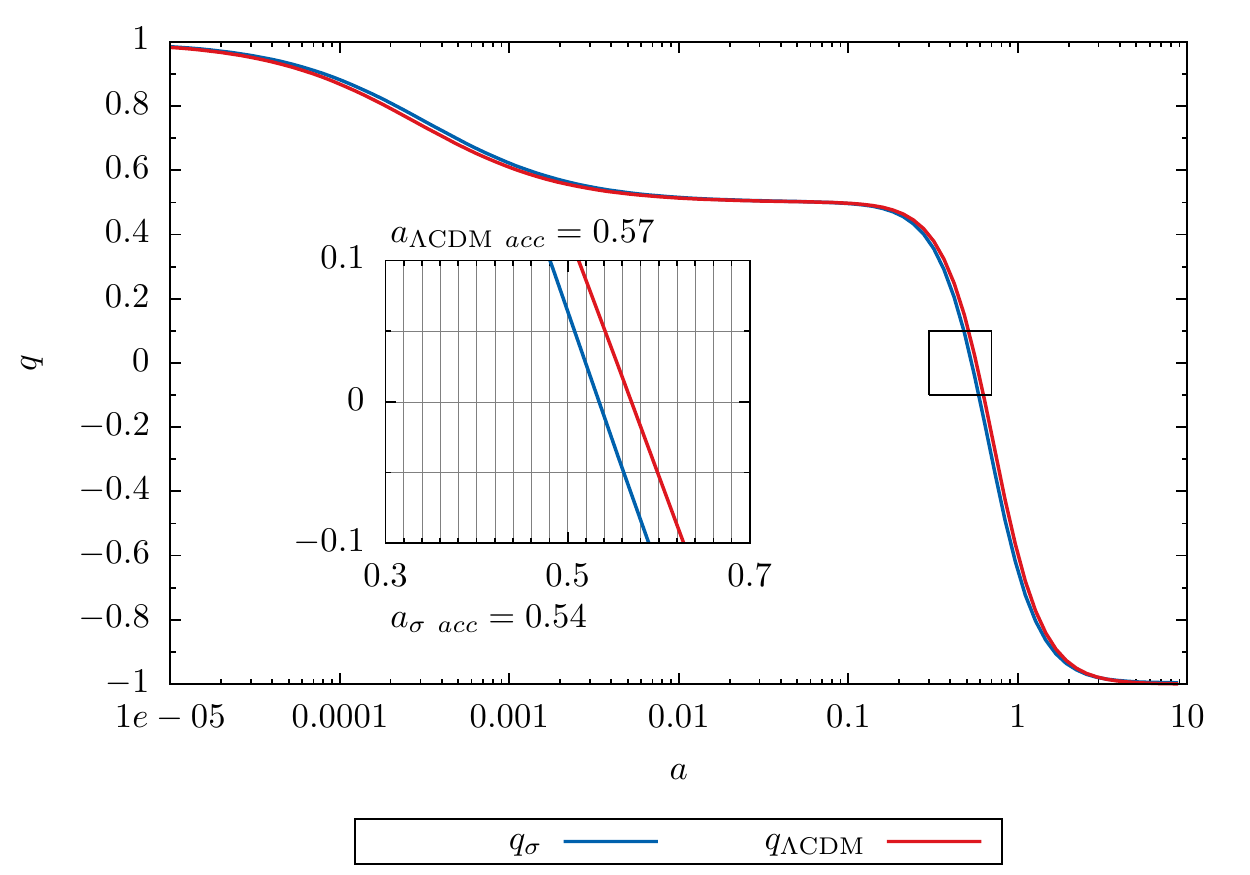}}
\caption{Evolution of the decelaration parameter in $\Lambda\mathrm{CDM}$ and $\sigma CDM$ models}
\label{ris:q}
\end{figure}

\begin{figure}[h]
\center{\includegraphics[width=1\linewidth]{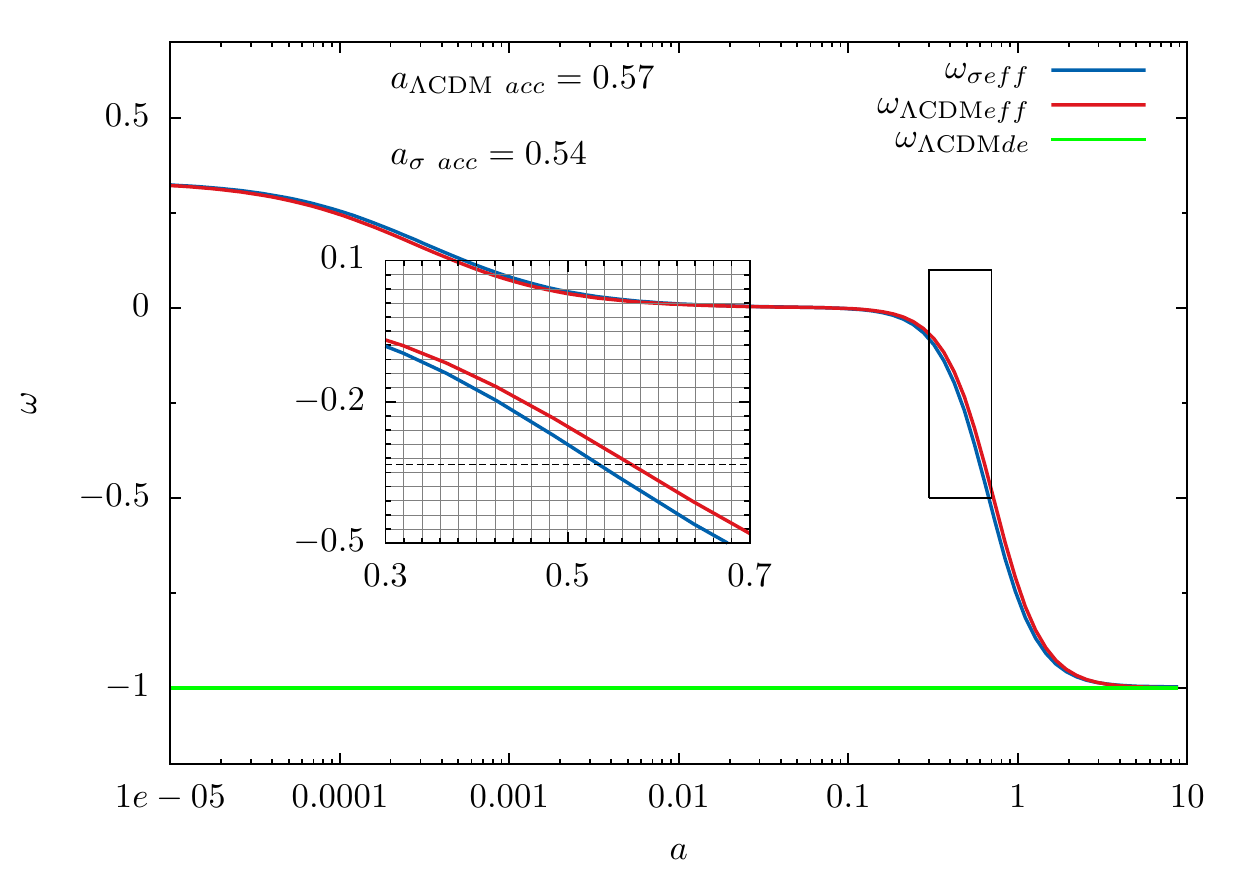}}
\caption{Evolution of the effective equation of state parameter in $\Lambda\mathrm{CDM}$ and $\sigma CDM$ models}
\label{ris:omega_eff}
\end{figure}

\begin{figure}[h]
\center{\includegraphics[width=1\linewidth]{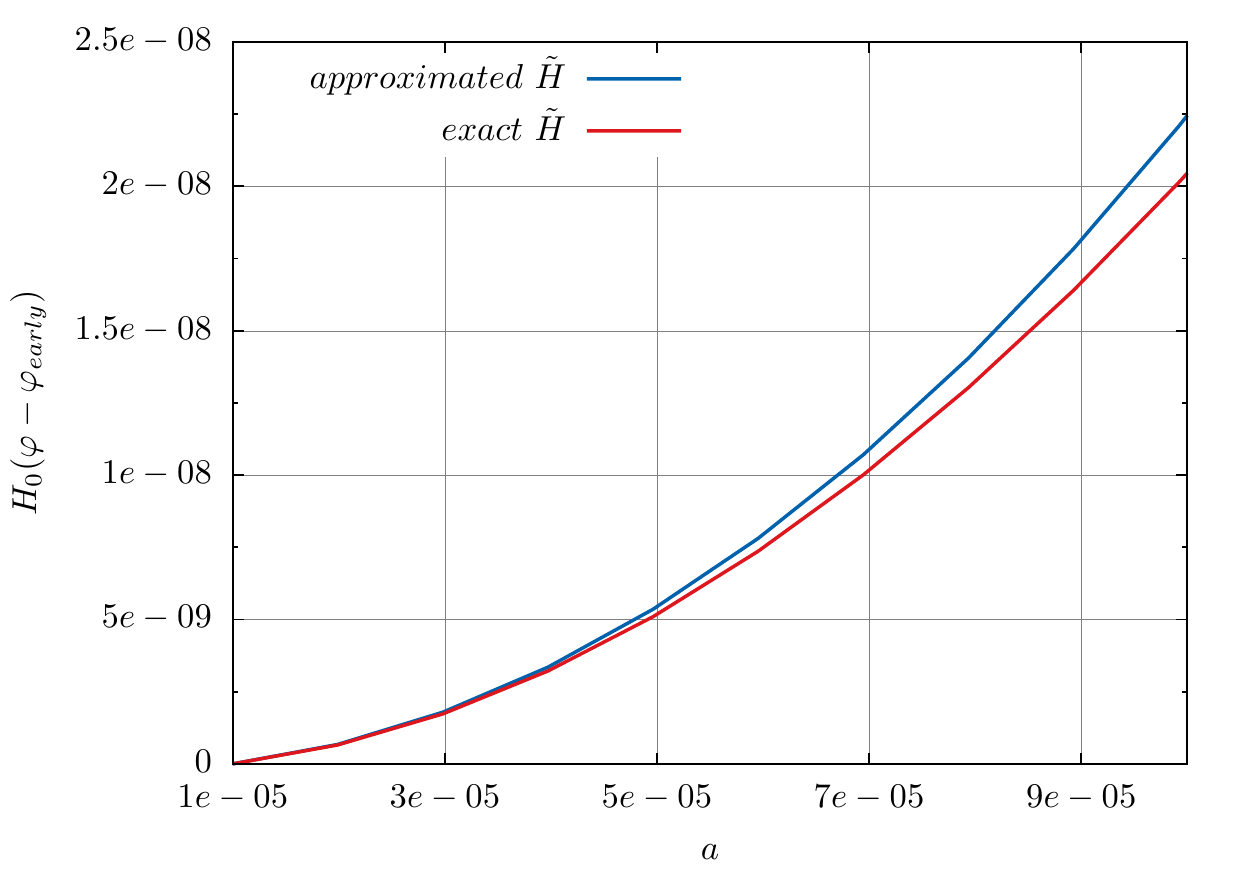}}
\caption{Early Universe approximation.}
\label{ris:early}
\end{figure}

\begin{figure}[h]
\center{\includegraphics[width=1\linewidth]{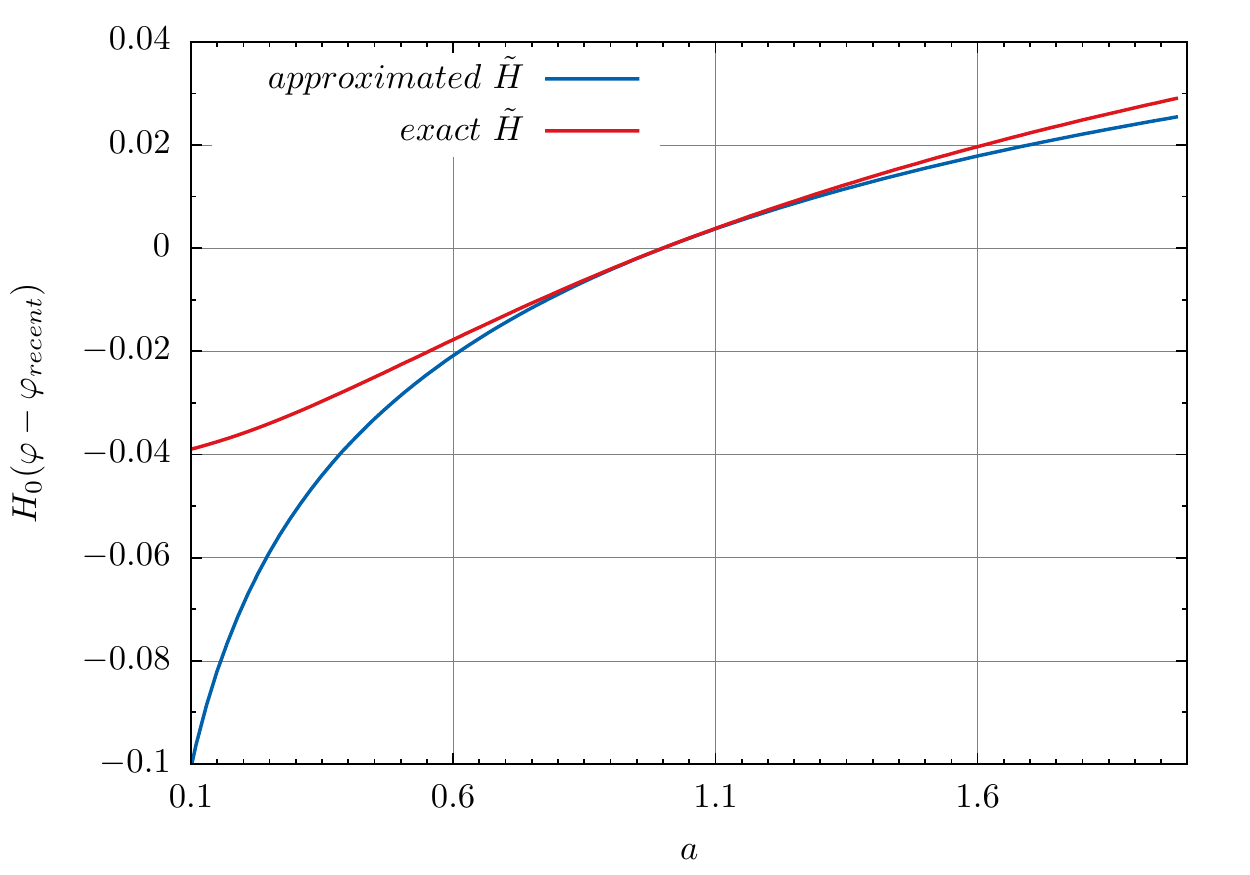}}
\caption{Recent Universe approximation.}
\label{ris:recent}
\end{figure}

\section{The reconstruction of the metric component $h_{22}$ and \\ the potential $V$}
\numberwithin{equation}{section}

In order to learn more about kinetic and potential interactions between DM and DE we must extend the standard reconstruction of the expansion history of the Universe \cite{sahni06} to a restoration of a functional
dependence on the scalar field $\ph$ for the target space metric component $h_{22}=h_{22}(\varphi)$
and the potential $V=V(\ph)$ of $\sigma$CDM.

Let us transform the ansatz (\ref{eq:f_g_ansatz-f}) (setting $h_{11}=1$)

\begin{equation}
  \frac{1}{2}\dot\ph^2=B=f,\quad\frac{1}{2}\left(\frac{d\ph}{dt}\right)^2=B,
\end{equation}
Changing variables from $t$ to $a$
\begin{equation}
  \frac{1}{2}\left(\frac{d\ph}{da}\right)^2=\left(\frac{dt}{da}\right)^2B,
\end{equation}
and introducing already known for us (\ref{eq:Hubble_sigma_DE}) Hubble parameter one can obtain
\begin{equation}
  \label{eq:ph_reconstr}
  \frac{1}{2}\ph'^2=\frac{B}{H_0^2a^2\tilde H^2}.
\end{equation}

Our next goal is to find the dependence $\ph=\ph(a)$, so we fix limits
of integration from some early epoch $a_i$ up to desired time moment,
corresponded to $a$
\begin{equation}
  \int^a_{a_i}H_0\ph'da=\sqrt{2}\int^a_{a_i}\frac{\sqrt{B} da}{a\tilde H}.
\end{equation}

The integral written here cannot be taken in an explicit form.
Nevertheless there is a possibility to use some approximations
based on a behavior of the different energy densities components
with a scale factor. Therefore following the idea of
\cite{Sur:2009jg} we consider
the early $a\ll1$ and recent Universe $a\approx1$ approximations.

Let us start from the case $a\ll1$. The Universe is known to be radiation dominated at the very early
times.  This means that all components contribution in Hubble
parameter is negligible in comparison with the radiation term.  So such
observation makes possible to do integration in

\begin{equation*}
  H_0\left(\ph(a)-\ph(a_i)\right)=
  \int^{a}_{a_{i}}\frac{\sqrt{2B}}{a\sqrt{\frac{\Omega_{r0}}{a^{4}}}}
da=\frac{\sqrt{B}}{\sqrt{2\Omega_{r0}}}\left({a^2-a^2_i}\right).
\end{equation*}

We can normalize the scalar field on today's critical density
$\ph\rightarrow\frac{\ph}{\sqrt{\rho_{c0}}}$, which gives us $\tilde
B$ instead of $B$ in (\ref{eq:ph_reconstr}).  Here
$\ph(a_i)=\ph(a=a_i)$, where $a_i$ is the fixed value of a scale
factor which will be taken equal to $10^{-5}$ and normalized to
current value $a_0$. It will be helpful for our analysis to introduce
$\tilde\ph_{early}=\ph(a_{i})-\frac{\sqrt{\tilde{B}}}
{H_{0}\sqrt{2\Omega_{r0}}}{a^2_i}= \ph(a_{i})+const$ and
$\ph_{early}=\ph(a_{i})$.

Now we have to invert $\ph=\ph(a)$ dependence
\begin{equation*}
  H_{0}\left( \ph-\tilde\ph_{early} \right)=\frac{\sqrt{\tilde{B}}}{\sqrt{2\Omega_{r0}}}{a^2}
\end{equation*}
to get the $a=a(\ph)$ dependence
\begin{equation*}
  a=\sqrt{\frac{H_0\sqrt{2\Omega_{r0}}}{\sqrt{\tilde{B}}}(\ph-\tilde\ph_{early})}.
\end{equation*}

If we know $a$ we can write down the chiral metric
component $h_{22}$ as a function on $\ph$

\begin{equation*}
  h_{22}=a^{-3}=\left(\frac{H_0\sqrt{2\Omega_{r0}}}{\sqrt{\tilde{B}}}\left( \ph-\tilde\ph_{early} \right)\right)^{-3/2}
\end{equation*}

Taking the constant $V_{0}=\Lambda-B$ in (\ref{eq:pot_scal}) one can obtain

\begin{equation*}
  V=V_{0}-3B\ln \left[ \frac{H_0\sqrt{2\Omega_{r0}}}{\sqrt{\tilde{B}}}\left( \ph-\tilde\ph_{early} \right) \right]+
   C \left[\frac{H_0\sqrt{2\Omega_{r0}}}{\sqrt{\tilde{B}}}\left( \ph-\tilde\ph_{early} \right)\right]^{-3/2}.
\end{equation*}

Thus we have finished the procedure of reconstruction of $h_{22}$ and $V$ for the
very early epoch of the Universe evolution when $a\ll1$.

Next step in consideration is the recent Universe approximation
 with $a\approx1$. Transforming the Hubble parameter to the form
~(\ref{eq:Hubble_sigma_DE})
\begin{equation*}
  \tilde H^2(a)=1+\frac{\Omega_{r0}}{a^{4}}(1-a^4)+\frac{\Omega_{m0}}{a^3}\left(1-a^3\right)-6\tilde{B}\ln a,
\end{equation*}
we obtain 
\begin{equation}
  \label{eq:taylor_rec_approx}
  \tilde H^2(a)=1+\frac{\Omega_{r0}}{(1-(1-a))^{4}}-\Omega_{r0}+\frac{\Omega_{0m}}{(1-(1-a))^3}-\Omega_{0m}-6\tilde{B}\ln(1-(1-a)).
\end{equation}
Let us apply the Taylor expansion about $(1-a)\approx0$ up to
first order terms in~(\ref{eq:taylor_rec_approx}). The result is
\begin{equation}
  \label{eq:rec_approx_Hubble}
  \tilde H^2(a)=1+\left(4\Omega_{r0}+3\Omega_{\sigma m0}+6\tilde{B}\right)(1-a),
  \quad\Omega_{\sigma m0}=\Omega_{b0}+\Omega_{\sigma cdm0}.
\end{equation}

From this moment we are able to fulfill the reconstruction procedure.
Dividing scalar field on $\sqrt{\rho_{c0}}$ once again
we come to
\begin{equation}\label{int-H0}
  H_0 \left( \ph(a)-\ph(a_{i}) \right)=\int^a_{a_{i}}\frac{\sqrt{2\tilde{B}}da}{a\sqrt{1+\left(3\Omega_{0m}+4\Omega_{r}+6\tilde{B}\right)(1-a)}}=
\int^a_{a_{i}}\frac{\sqrt{2\tilde{B}}da}{a\sqrt{\alpha-\beta a}},
\end{equation}
where
\begin{equation*}
  \beta=3\Omega_{m0}+4\Omega_{r0}+6\tilde{B}=3\cdot0.23+4\cdot5\cdot10^{-5}\cdot(1+0.6)+6\cdot0.007>0,
\end{equation*}
\begin{equation*}
  \alpha=1+\beta>1.
\end{equation*}

It is essential for the subsequent analysis that the best--fit value of
$\tilde{B}$ is known, so we have an opportunity to make calculation
of the integral (\ref{int-H0}). Such a type of an integral is calculated by
\begin{equation*}
  \int\frac{dx}{\sqrt{x}(x-b)}=-\frac{1}{\sqrt{b}}\ln\left|\frac{\sqrt{x}+\sqrt{b}}{\sqrt{x}-\sqrt{b}}\right|,~b>0.
\end{equation*}


One more issue is about a transition through $a=1$. This scale factor value should be
explicitly presented in the expression for $\ph$
\begin{equation}
  \label{eq:rec_univ_split}
  H_0\left(\ph-\ph(a=a_i)\right)=\sqrt{2\tilde{B}}\left\{\int_{a_i}^1\frac{d a}{ a\tilde H}+\int_{1}^a\frac{d a}{a\tilde H}\right\}.
\end{equation}

With the help of

\begin{equation*}
  \tilde{\ph}_{recent}=\ph(a=a_{i})+\frac{\sqrt{2\tilde{B}}}{H_{0}}\frac{1}{\sqrt{\alpha}}
\ln \left| \frac{\sqrt{\alpha-\beta a_i}+\sqrt{\alpha}}{\sqrt{\alpha-\beta a_i}-\sqrt{\alpha}} \right|,
\end{equation*}
and
\begin{equation*}
  \ph_{recent}=\ph(a=a_{i})+\frac{\sqrt{2\tilde{B}}}{H_{0}}\frac{1}{\sqrt{\alpha}} \left[
-\ln \left| \frac{\sqrt{\alpha-\beta}+\sqrt{\alpha}}{\sqrt{\alpha-\beta}-\sqrt{\alpha}} \right|
+\ln \left| \frac{\sqrt{\alpha-\beta a_i}+\sqrt{\alpha}}{\sqrt{\alpha-\beta a_i}-\sqrt{\alpha}} \right|\right],
\end{equation*}
one can carry out computaions further in more compact form.
Plausibility of the early and recent approximations can be deduced from
the comparison of $H_0(\ph-\ph_{early})$ and $H_0(\ph-\ph_{recent})$
for the exact~(\ref{eq:Hubble_sigma_DE}) and
approximate~(\ref{eq:rec_approx_Hubble}) Hubble parameter expressions and the
time limits (for which corresponding approximations are hold on)
will be extracted graphically.

The reconstruction is performed as usual when $a=a(\ph)$ is obtained. Using notation of the
$\tilde{\ph}_{recent}$ in (\ref{eq:rec_univ_split})
we have
\begin{equation*} H_0
\left(\ph-\tilde{\ph}_{recent}\right)=-\frac{\sqrt{2\tilde{B}}}{\sqrt{\alpha}}\ln\Bigl|\frac{\sqrt{\alpha-\beta
a}+\sqrt{\alpha}}{\sqrt{\alpha-\beta a}-\sqrt{\alpha}}\Bigr|,
\end{equation*}
and

\begin{equation*}
  a=\frac{\alpha}{\left( \alpha-1 \right)\cosh^2 \left(
A(\ph)\right)},\quad A(\ph)=-\frac{\sqrt{\alpha}}{2\sqrt{2\tilde{B}}} H_{0}\left(
\ph-\tilde{\ph}_{recent} \right).
\end{equation*}

We can substitiute this result to 
$h_{22}$~(\ref{eq:h22}) and $V$~(\ref{eq:pot_scal}) to get
\begin{equation*}
  h_{22}=\left( \frac{\alpha-1}{\alpha} \right)^3\cosh^{-6} \left(
A(\ph)\right),
\end{equation*}

\begin{align*}
  V=V_0-6B\ln \left[ \frac{\alpha}{\left( \alpha-1 \right)\cosh^2 \left(
A(\ph)\right)} \right]+C\left[\frac{\alpha}{\left( \alpha-1 \right)\cosh^2 \left(
A(\ph)\right)}\right]^{-3}.\nonumber
\end{align*}

\section{Background dynamics of the model}
\numberwithin{equation}{section}

One of the most important cosmological parameter used for the description of a background
evolution is a contribution to a critical density of the Universe.
The last is defined as $\Omega=\frac{\rho}{\rho_c}$.
For the chiral fields sector we have

\begin{equation*}
  \Omega_{\sigma}=\frac{\rho_{\sigma}}{\rho_c}=\frac{\rho_{\sigma}}{\tfrac{3H_{0}^2\tilde{H}^2}{8\pi G}}.
\end{equation*}
Using (\ref{eq:rho_sigma}) and
(\ref{eq:Hubble_sigma_DE}) one can obtain
\begin{equation*} \Omega_{\sigma}=\frac{\tilde{\Lambda}-6\tilde{B}\ln
a+2\tilde{C}a^{-3}}{\Omega_{\sigma\Lambda0}+\Omega_{\sigma
cdm0}a^{-3}+\Omega_{b0}a^{-3}+\Omega_{r0}a^{-4}-6\tilde{B}\ln a}.
\end{equation*}
In order to understand a general picture of the Universe
evolution and to analyze periods of domination by various species of the Universe we
represent the residual components of $\sigma$CDM model

\begin{equation*}
  \Omega_{\sigma r}=\frac{\Omega_{r0}a^{-4}}{\tilde{H}^2},\quad
  \Omega_{\sigma b}=\frac{\Omega_{b0}a^{-3}}{\tilde{H}^2},\quad
  \Omega_{\sigma m}=\frac{\Omega_{\sigma m0}a^{-3}}{\tilde{H}^2},\quad
  \Omega_{\sigma
de}=\frac{\Omega_{\sigma\Lambda0}-6\tilde{B}\ln
a}{\tilde{H}^2},
\end{equation*}
where $\tilde{H}$ comes from (\ref{eq:Hubble_sigma_DE}).
The latter quantity is responsible for the late accelerated expansion of
the Universe, supported by $\sigma$CDM model.

In the $\Lambda$CDM we have
\begin{equation*}
\Omega_{r}=\frac{\Omega_{r0}a^{-4}}{\tilde{H}^{2}},\quad
\Omega_{b}=\frac{\Omega_{b0}a^{-3}}{\tilde{H}^{2}},\quad
\Omega_{m}=\frac{\Omega_{m0}a^{-3}}{\tilde{H}^{2}},\quad
\Omega_{de}=\frac{\Omega_{\Lambda}}{\tilde{H}^{2}}.
\end{equation*}

Here $\tilde{H}^{2}$ is given by
\begin{equation}
  \label{eq:lcdm_tilde_H}
  \tilde{H}^{2}=\Omega_{\Lambda0}+\Omega_{m0}a^{-3}+\Omega_{b0}a^{-3}+\Omega_{r0}a^{-4}.
\end{equation}

Let us turn our attention to the effective equation of the state
parameter

\[\omega_{eff}=\frac{\sum_{\alpha}p_{\alpha}}{\sum_{\alpha}\rho_{\alpha}}.\]

It is necessary to take into account expressions for densities and
pressures of the chiral
fields~(\ref{eq:rho_sigma}),~(\ref{eq:press_sigma}) and the other
components, together with~(\ref{eq:Hubble_sigma_DE})

\begin{equation*}
  \rho_{r}=\rho_{r0}a^{-4}=
\Omega_{r0}\rho_{c0}a^{-4},\quad\rho_{b}=\rho_{b0}a^{-3}
=\Omega_{b0}\rho_{c0}a^{-3},\quad p_{r}=\frac{1}{3}\rho_{r},\quad p_{b}=0.
\end{equation*}
Then in the $\sigma$CDM model we will have
\begin{equation*}
  \omega_{\sigma(eff)}=\frac{p_r+p_b+p_{\sigma}}{\rho_r+\rho_b+\rho_{\sigma}}
  =\frac{\frac{1}{3}\Omega_{r0}a^{-4}+\left(-\Omega_{\sigma\Lambda0}+
6\tilde{B}\ln a+2\tilde{B}\right)}
{\Omega_{r0}a^{-4}+\Omega_{\sigma m0}a^{-3}+\Omega_{\sigma\Lambda0}-6\tilde{B}\ln a}.
\end{equation*}
At the same time for $\Lambda$CDM model the effective equation of state parameter is
\begin{equation*}
  \omega_{\Lambda\mathrm{CDM}(eff)}=\frac{1}{2}\frac{2\Omega_{r0}a^{-4}+\Omega_{m0}a^{-3}+(-2\Omega_{\Lambda0})}
{\Omega_{\Lambda0}+\Omega_{m0}a^{-3}+\Omega_{b0}a^{-3}+\Omega_{r0}a^{-4}}.
\end{equation*}

Using the definition of the deceleration parameter $q$ broadly used for background dynamics studies we can obtain
\begin{equation*}
  q=-\frac{\ddot a a}{\dot a^2}=\frac{\frac{4\pi G}{3}
    \left(\sum_\alpha\rho_\alpha+3p_\alpha\right)}{\frac{8\pi G}{3}{\sum_\alpha\rho_\alpha}}.
\end{equation*}

Presence in the $q$ the second derivative of the scale factor gives
us evidence of the transition from decelaration to acceleration
epoch at the time when $q=0$. The deceleration parameter for chiral sector takes the view

\begin{equation*} q_{\sigma}=\frac{1}{2}\frac{\Omega_{\sigma
m0}a^{-3}+2\Omega_{r0}a^{-4}+12\tilde B\ln
a-2\Omega_{\sigma\Lambda0}+6\tilde{B}}{\tilde H^2},
\end{equation*}
where $\tilde{H}^{2}$ is defined in (\ref{eq:Hubble_sigma_DE}).
For $\Lambda$CDM model the expression for deceleration parameter looks like
\begin{equation*}
q_{\Lambda\mathrm{CDM}}=\frac{1}{2}\frac{\left(\Omega_{b0}
a^{-3}+2\Omega_{r0}a^{-4}+\Omega_{cdm0}a^{-3}-2\Omega_{\Lambda0}\right)}{\tilde
H^2}
\end{equation*}
with Hubble parameter taken from (\ref{eq:lcdm_tilde_H}).
The evident differences both in numerator and denominator in $q_{\sigma}$ and
$q_{\Lambda\text{CDM}}$ inevitably lead to distinctive evolution
of the Universe if it is supported by $\sigma$CDM or $\Lambda$CDM models.

It is known feature of $\omega_{eff}$ that a moment of time of
deceleration/accelaration transition corresponds to value $-1/3$
crossing. It is also well--known result this time to be exactly
equivalent to those obtained from $q$ analysis. We would like to
point out here that in $\Lambda$CDM model equation of state of
the dark energy parameter is equal
$\omega_{\Lambda\mathrm{CDM}de}=-1$.

\section{Discussion}
\numberwithin{equation}{section}

Fig.\ref{ris:mu} shows good agreement of supernovae data with
$\sigma$CDM model taken with the best--fit parameters values.  This
fact confirm the validity of proposed model, i.e., $\sigma $CDM model
does not contradict to observational data and may serve as a good
dynamical alternative to $\Lambda$CDM.

In fig.  \ref{ris:contours} the confidence contours are depicted. We
keep only positive values of parameter $\tilde{B}$ in order to prevent
a crossing of the phantom divide.  

One can see from the evolution of the individual densities $\Omega_i$,
deceleration parameters $q$ and effective equation of state parameters
$\omega_{eff}$ (figs. \ref{ris:Omega}, \ref{ris:q} and
\ref{ris:omega_eff}) that accelerated expansion takes place earlier in
the Universe supported by $\sigma$CDM. Also one may notice that
radiation/matter domination transition occurs earlier in $\Lambda$CDM
model. These observations are in the full agreement with a smaller
total matter amount including cold dark and baryonic components in
$\sigma$CDM model in comparison to $\Lambda$CDM model.

The graphical comparison (see fig. \ref{ris:q} and
fig. \ref{ris:omega_eff}) of the $a_{\Lambda\mathrm{CDM} acc}$ and
$a_{\sigma acc}$ (taken from the scale factor values corresponding to
$-1/3$ and $0$ crossing) gives us clear evidence for equality of the
transitions to accelerate expansion in corresponding models. This
observation is concluded from $q$ and $\omega_{eff}$ values and has
been already mentioned above.

From fig. \ref{ris:early} one can conclude that the early Universe
approximation holds for $a=10^{-5}$ up $a=5\cdot 10^{-5}$ scale factor
values. The recent Universe approximation depicted on fig.~\ref{ris:recent}
is true from $a=0.8$ to $a=1.2$ values. Let us remind that validity of
the early and recent approximations comes from confrontation of
$H_0(\ph - \ph_{early})$ and $H_0(\ph - \ph_{recent})$.
The deviation for approximated and exact $\tilde{H}$ is associated
with the lost of domination of DE for the early times.

In conclusion it needs to stress that we first time reconstructed from 
observations the kinetic interaction between DM and DE in the form of chiral metric component $h_{22}$ for $\sigma$CDM. Also we have hope that the reconstruction techniques presented here may be useful for exact solution construction because of obtaining $h_{22}$ from observational data.

\section*{Acknowledgments}
SVC is thankful to the University of
KwaZulu-Natal, the University of Zululand and the NRF for financial
support and warm hospitality during his visit in 2012 to South Africa
where the part of the work was done. RRA is grateful to participants
of the scientific seminars headed by Melnikov V.N.(Institute of
Gravitation and Cosmology, Moscow), Rybakov Yu.P. (PFUR, Moscow) and
Sushkov S.V. (KFU, Kazan) for valuable comments and criticize.


\begin{thebibliography}{10}

  \bibitem{Suzuki:2011hu}
N.~Suzuki, D.~Rubin, C.~Lidman, G.~Aldering, R.~Amanullah {\em et~al.}, {\em
  Astrophys.J.} {\bf 746}, 85 (2012).

\bibitem{Percival:2009xn}
W.~J. Percival {\em et~al.}, {\em Mon.Not.Roy.Astron.Soc.} {\bf 401}, 2148 (2010).

\bibitem{Komatsu:2010fb}
E.~Komatsu {\em et~al.}, {\em Astrophys.J.Suppl.} {\bf 192}, 18 (2011).

\bibitem{tsuji-10}
S.~Tsujikawa, astro-ph/1004.1493, (2010).

\bibitem{cosats06}
E.~J. Copeland, M.~Sami and S.~Tsujikawa, {\em Int.J.Mod.Phys.} {\bf D15}, 1753 (2006).

\bibitem{DarkEnergy1103}
M.~Li, X.-D. Li, S.~Wang and Y.~Wang, {\em Commun.Theor.Phys.} {\bf 56}, 525 (2011).

\bibitem{Padmanabhan}
T.~Padmanabhan, {\em Phys.Rept.} {\bf 380}, 235 (2003).

\bibitem{QuintomReview}
Y.-F. Cai, E.~N. Saridakis, M.~R. Setare and J.-Q. Xia, {\em Phys.Rept.} {\bf
  493}, 1 (2010).

\bibitem{chervon12qm}
Chervon S. V., {\em Quantum Matter.} {\bf 2}, 1 (2013).

\bibitem{ch97gc}
Chervon S. V., {\em Grav.Cosmol.} {\bf 3}, 145 (1997).

\bibitem{ch02gc}
Chervon S. V., {\em Grav.Cosmol.} {\bf 3}, 32 (2002).

\bibitem{brochesush}
K.~Bronnikov, S.~Chervon and S.~Sushkov, {\em Grav.Cosmol.} {\bf 15}, 241 (2009).

\bibitem{bcmk12qm}
Beesham A., Chervon S. V., Maharaj S. D., Kubasov A. S., {\em Quantum
Matter.} {\bf 2}, (2013).

\bibitem{Beesham}
A.~Beesham, S.~Chervon and S.~Maharaj, {\em Class.Quant.Grav.} {\bf 26}, 075017 (2009).

\bibitem {chzhsh97plb}
Chervon S. V. , Zhuravlev V. M. and Shchigolev V. K., Phys. Lett.
{\bf
B 398}, 269 (1997).


\bibitem{chekos2003}
S.~V.~Chervon, N.~A.~Koshelev, {\em Grav.Cosmol.} {\bf 9}, 196
(2003).

\bibitem{panche11}
Panina O. G., Chervon S. V. in {\it On the pre-inflationary dark
sector fields influence on the cosmological perturbations}, Proc.
of Sci. The XXth International Workshop HEP and QFT, Sept.-24 --
Oct.1, 2011; http://pos.sissa.it, (2011).


\bibitem{arrche12gc}
R.~R.~Abbyazov, S.~V.~Chervon, {\em Grav.Cosmol.} {\bf 18},
262 (2012).

\bibitem{Sur:2009jg}
S.~Sur, astro-ph/0902.1186, (2009).

\bibitem{Generalizationquintom}
L.~P. Chimento, M.~I. Forte, R.~Lazkoz and M.~G. Richarte, {\em Phys.Rev.} {\bf
 D79}, 043502 (2009).

\bibitem{Quintommixed}
E.~N. Saridakis and J.~M. Weller, {\em Phys.Rev.} {\bf D81}, 123523 (2010).

\bibitem{twoscalarfields}
C.~van~de Bruck and J.~M. Weller, {\em Phys.Rev.} {\bf D80}, 123014 (2009).

\bibitem{Linde90}
A. D. Linde, {\it Particle Physics and Inflationary Cosmology}
Harwood Acad. Publ., Paris--New York, (1990) [Russ. original,
Nauka, Moscow, 1990].


\bibitem{chervon95aa}
S.~V.~Chervon {\em J. Astrophys. Astron., Suppl.}, {\bf 16},
65 (1995).


\bibitem{chepan2010}
S.~Chervon and O.~Panina, {\em Journal "Vestnik RUDN"} \textbf{4}, 121 
(2010).

\bibitem{chepan2008}
S.~Chervon and O.~Panina, {\em Vestnik SamGU, Estestvennonauchnaya
seriya} {\bf No.8/1(67)}, 611 (2008).

\bibitem{Garcia-Bellido}
Garcia-Bellido J., astro-ph/0502139, (2005).


\bibitem{Perlmutter}
S.~Perlmutter {\em et~al.}, {\em Astrophys.J.} {\bf 517}, 565 (1999).

\bibitem{Riess}
A.~G. Riess {\em et~al.}, {\em Astron.J.} {\bf 116}, 1009 (1998).

\bibitem{Comparisondarkenergy}
M.~Li, X.~Li and X.~Zhang, {\em Sci.China Phys.Mech.Astron.} {\bf 53}, 1631 (2010).

\bibitem{Hu:1995en}
W.~Hu and N.~Sugiyama, {\em Astrophys.J.} {\bf 471}, 542 (1996).

\bibitem{sahni06}
Sahni V., Starobinsky A., Int.J.Mod.Phys. {\bf D15}, 2105 (2006).

\bibitem{ch95iv}
S.~Chervon, {\em Russ.Phys.J.} {\bf 38}, 539 (1995).

\bibitem{arrche13vr}
R.~R.~Abbyazov, S.~V.~Chervon, { accepted for publication in \em Journal "Vestnik RUDN"} (2013).

\end{thebibliography}

\end{document}